\documentstyle[preprint,aps]{revtex}

\begin{document}

\draft
\preprint{}
\title{Hidden Local Symmetry and Effective Chiral Theory \\ for Vector and Axial-vector Mesons}
\author{ B.A. Li$^{\dagger}$ and Y.L.  Wu$^{ \ddagger}$  }
\address{
$\dagger$\ Department of Physics, University of Kentucky, Lexington, KY 40506, USA\\ 
 $\ddagger$\ Institute of Theoretical Physics, Academia Sinica, \\
 P.O. Box 2735, Beijing 100080, P.R. China}
\date{}
\maketitle

\begin{abstract}
 We present the effective chiral Lagrangian of mesons (peusodoscalars, vectors and axial-vectors)
 obtained in the chiral limit by using two approaches. The first approach is based on symmetries: 
the explicit global
 chiral symmetry and hidden local chiral symmetry. In this approach, it is noticed that
 there are in general fourteen interacting terms up to the dimension-four of covariant derivative
 for meson fields rather than the usual eleven interacting terms given in literature from
 hidden local symmetry approach. Of particular, the additional terms are found to be very important for understanding
 the vector meson dominance and providing consistent predictions on the decay rates of $a_1\rightarrow \gamma \pi$ and
 $a_1\rightarrow \rho \pi$ as well as for resulting a consistent effective chiral Lagragian with chiral perturbation theory.
 The second approach is motivated from the chiral symmetry of chiral quarks and
 the bound state solutions of nonperturbative QCD at low energy and large $N_c$. The second approach is more
 fundamental in the sense that it is based on the QCD Lagrangain of quarks and only relies on
 two basic parameters in addition to the ones in the standard model. As a consequence, it allows us to extract,
 in terms of only two basic parameters, all the fourteen parameters in the more general effective
 Lagrangian constructed from symmetries in the first approach. It is surprising to note that except the necessity of
 three additional new interacting terms introduced in this paper, the resulting values of the coupling constants
 for other three interacting terms at the dimension-four are also quite different from the ones
 given in the literature. It is likely that the structures of
 the effective chiral Lagrangian for the dimension-four given in the literatures by using hidden local symmetry
 are incomplete and consequently the resulting coulpings are not reliable. It is shown that the more
 general effective chiral Lagrangian given in the present paper shall provide a more consistent prediction for all the low
 energy phenomenology of $\rho-a_1$ system and result in a more consistent description on the low energy
 behavior of light flavor mesons. Its fourteen parameters up to the dimension-four of covariant derivative
 may be uniquely determined from the effective chiral theory based on the second approach, which is consistent with the
 chiral perturbation theory.
\end{abstract}

\newpage

\section{Introduction}

 QCD with gauge symmetry SU(3) has been introduced to describe the strong interactions among quarks.
 Its high energy behavior, in particular, the asymptotic behavior, has been succesfully characterized
 by perturbative QCD. While the low energy dynamics of QCD remains unsolved due to the nonperturbative
 effects of strong interactions. Actually, QCD was motivated from the studies of low energy dynamics of
 hadrons. A lot of useful and practical approaches, such as current algebra, PCAC, sum rules, dispersion analysis,
 vector meson dominance, Green function approach, Nambu-Jona-Lasinio model, chiral quark model, {\it etc.},
 have been adopted to study the
 low energy behavior of hadronic physics, many of them were developed even before QCD was discovered.
 It is now believed that hadrons are bound states of quarks and gluons as the solutions of
 nonperturbative QCD.  It also becomes clear that the quark confinement and chiral symmetry breaking
 result from nonperturbative QCD effects. The success of current algebra\cite{CM} with PCAC
 is mainly because it reflects the (approximate) chiral invariance of the QCD lagrangian. Base on the successful
 descriptions of these approaches, it is then natural to study an effective theory based on
 chiral quarks with (approximate) chiral symmetry and bound state solutions of
 nonperturbation QCD. Though the meson physics has been widely studied for a long time, theoretically,
 there remain some open questions: what is a more realistic effective Lagrangian that can systematically
 describe the pseudoscalars, vector and axial-vector mesons, and how to establish such an effective Lagrangian
 with less parameters. In this paper, we will issue such questions along two lines. Firstly, we adopt the chiral
 symmetry to construct more general effective Lagrangain. It is based on the fact that the QCD lagrangian
 possesses a global chiral symmetry $U(3)_L\times U(3)_R$ in the limit of zero light qurak mass $m_q\rightarrow 0$.
 The vector and axial-vector mesons are treated as dynamical gauge bosons of hidden local chiral symmetry
 $\hat{U}$(3)$_L$ $\times$ $\hat{U}$(3)$_R$. This approach has widely been adopted in the literature and several
 review articles\cite{BKY,UM} have extensively discussed such an approach. A simple but complete description
 on this approach will be presented in the section II. In particular, we will see that up to the dimension-four
 of meson fields, the resulting effective Lagrangian has in general fourteen interacting terms rather than
 eleven interacting terms given in the literature with hidden local symmetry approach\cite{BFY,BKY,UM}.
 Three important interacting terms at the dimension-four have actually been missed in
 the mentioned literature. It is those additional terms that cancel the strong momentum dependence of the
 $\rho-\pi-\pi$ coupling $f_{\rho\pi\pi}$, and also it is those terms that ensure the $\rho-$ meson
 dominance in $a_1\rightarrow \gamma \pi$ decay and result in consistent prediction on the decay rates
 of $a_1\rightarrow \gamma \pi$ and $a_1\rightarrow \rho \pi$. In section III, we present an effective Lagrangian
 derived from the QCD Lagrangian of chiral quarks with (approximate) global chiral symmetry
 and bound state solutions of nonperturbative QCD. Where only two parameters are introduced in addtion
 to the ones in the standard model. Such an effective chiral theory (ECT) has been shown to provide
 a lot of consistent results on low energy phenomena for the vector and axial-vector meson system\cite{BAL1,BAL2}.
 In section IV, it is shown that with an appropriate gauge fixing of hidden local chiral symmetry,
 fourteen parameters appearing in the more general effective Lagarangian construced
 based on the explict global chiral symmetry and hidden local
 chiral symmetry can be uniquely extracted when comparing it with the effective Lagrangian of ECT. As a consequence,
 all low energy phenomenology of $\rho-a_1$ system, such as universality of the $\rho$-meson coupling,
 vector meson dominance, the $\rho-\pi-\pi$ coupling $f_{\rho\pi\pi}$, the KSFR relation
 $m_{\rho}^2 = f_{\rho\pi\pi}^2 f^2_{\pi}/2$, etc. can be
 consistently understood. While the Weinberg's sum rule is modified to be $g_{a}^2/(1-1/2\pi^2g^2) = g_{\rho}^2$.
 Our conclusions and remarks are given in the last section.

\section{Vectors, Axial-vectors and Hidden Local Symmetry}

   Neglecting the smallness of the light quark mass $m_q$ ($q=u,d,s$)
   in comparing with the chiral symmetry breaking scale, the strong interactions of
   light flavors possess a global chiral symmetry U(3)$_L$ $\times$ U(3)$_R$.
   The chiral symmetry is supposed to be spontaneously broken via the dynamical mechanism
   due to attractive gauge interactions, namely the chiral condensates $<\bar{q}q> $
   exist and lead to the Goldstone-like pseudoscalar mesons $\pi$, $K$,
   $\eta$. If the vector and axial-vector mesons are introduced as gauge bosons, they
   should not be corresponding to the gauge bosons via gauging the above global
   chiral symmetry U(3)$_L$ $\times$ U(3)$_R$, otherwise there exist,
   according to the Higgs mechanism, no independent degrees of freedoms for
   the Goldstone-like pseudoscalar mesons. On the other hand, the chiral gauge boson
   couplings to the light quarks must be invariant under the transformation of
   the global chiral symmetry U(3)$_L$ $\times$ U(3)$_R$ as the original QCD theory
   does. It is then motivated to introduce hidden local chiral symmetry
   $\hat{U}$(3)$_L$ $\times$ $\hat{U}$(3)$_R$ associated with the chiral gauge bosons
   $\hat{A}_L$ and $\hat{A}_R$.
   After the spontaneous breaking of the global chiral symmetry U(3)$_L$ $\times$ U(3)$_R$,
   the Goldstone-like pseudoscalar mesons are generated,
   the chiral gauge bosons associated with the hidden local gauge symmetry also
   turn out to be the vector and axial-vector mesons via an appropriate choice of the gauge transformation of
    the hidden local chiral symmetry $\hat{U}$(3)$_L$ $\times$ $\hat{U}$(3)$_R$. Such a gauge choice breaks the
    hidden local chiral symmetry and generates the masses of the vector and axial-vector mesons. In this note,
    we are limited to consider the part with chiral symmetry at zero quark mass and will not discuss the gauge anormalous part.

    Let us begin with introducing the necessary fields for constructing the chiral Lagrangian which is invariant under
    the global chiral symmetry U(3)$_L$ $\times$ U(3)$_R$.
    The chiral Lagrangian is supposed to describe the Goldstone-like pseudoscalars,
    both vector and axial-vector mesons which arise from the gauge bosons of
    the local chiral symmetry $\hat{U}$(3)$_L$ $\times$ $\hat{U}$(3)$_R$.

    The needed fields contain the nonlinear chiral fields: $\hat{\xi}_L(x)\in U(3)_L\times \hat{U}(3)_L$
    and $\hat{\xi}_R(x) \in  U(3)_R\times \hat{U}(3)_R$, which transform as
    \begin{eqnarray}
    & &  \hat{\xi}_L(x) \rightarrow g_L\ \hat{\xi}_L(x)\ G_L^{\dagger}(x) \ ; \qquad  g_L \in U(3)_L\ ,
    \quad G_L(x) \in  \hat{U}(3)_L  \\
    & &  \hat{\xi}_R(x) \rightarrow g_R\ \hat{\xi}_R(x)\ G_R^{\dagger}(x) \ ; \qquad  g_R \in U(3)_R\ ,
    \quad G_R(x) \in \hat{U}(3)_R
   \end{eqnarray}
    and the nonlinear field $\xi_M \in \hat{U}(3)_L\times \hat{U}(3)_R$ tranforming as
    \begin{equation}
    \xi_M(x) \rightarrow G_L(x)\ \xi_M(x)\ G_R^{\dagger}(x) \ ;
    \qquad  (G_L(x),\ G_R(x)) \in \hat{U}(3)_L\times \hat{U}(3)_R
    \end{equation}
    The chiral gauge fields $\hat{A}_L$ and $\hat{A}_R$ corresponding to the local chiral symmetry
    $\hat{U}$(3)$_L$ $\times$ $\hat{U}$(3)$_R$ transform as
     \begin{eqnarray}
    & &  \hat{A}_L(x) \rightarrow G_{L}(x) ( \hat{A}_L(x) + i\partial ) G_L^{\dagger}(x), \\
    & &  \hat{A}_R(x) \rightarrow G_{R}(x) ( \hat{A}_R(x) + i\partial ) G_R^{\dagger}(x)
   \end{eqnarray}
   With the above nonlinear chiral fields, one can construct the nonlinear field $U(x) \in U(3)_L \times U(3)_R$
   transforming only under the global chiral symmetry action
     \begin{eqnarray}
    & & U(x)\equiv \hat{\xi}_L(x)\xi_M(x)\hat{\xi}_R^{\dagger}(x),\\
    & &  U(x) \rightarrow g_L\ U(x)\ g_R^{\dagger} \ ; \qquad  (g_L, \ g_R) \in U(3)_L\times U(3)_R
    \end{eqnarray}
   Similarly, we can construct the chiral gauge bosons
    \begin{eqnarray}
    & &  a_L(x) = \hat{\xi}_{L}(x) ( \hat{A}_L(x) + i\partial ) \hat{\xi}_L^{\dagger}(x)\equiv
    \hat{\xi}_{L}(x) iD \hat{\xi}_L^{\dagger}(x), \\
    & &  a_R(x) = \hat{\xi}_{R}(x) ( \hat{A}_R(x) + i\partial ) \hat{\xi}_R^{\dagger}(x) \equiv
    \hat{\xi}_{R}(x) iD \hat{\xi}_R^{\dagger}(x)
   \end{eqnarray}
   which transform under the global chiral symmetry as
    \begin{eqnarray}
     a_L(x) \rightarrow g_{L}\ a_L(x)\ g_L^{\dagger}, \qquad a_R(x) \rightarrow g_{R}\ a_R(x)\ g_R^{\dagger}
   \end{eqnarray}
   The gauge field strengths of the local chiral symmetry are defined as
   \begin{eqnarray}
    & &  \hat{F}_L^{\mu\nu}(x) = \partial^{\mu}\hat{A}_L^{\nu} - \partial^{\nu}\hat{A}_L^{\mu}
    -i [ \hat{A}_L^{\mu}, \hat{A}_L^{\nu}] \nonumber \\
    & &  \hat{F}_R^{\mu\nu}(x) = \partial^{\mu}\hat{A}_R^{\nu} - \partial^{\nu}\hat{A}_R^{\mu}
    -i [ \hat{A}_R^{\mu}, \hat{A}_R^{\nu}]
   \end{eqnarray}
   and the field strengths of the chiral gauge bosons corresponding to the global chiral symmetry are
   given by
   \begin{eqnarray}
    & &  F_L^{\mu\nu}(x) = \partial^{\mu}a_L^{\nu} - \partial^{\nu}a_L^{\mu} -i [ a_L^{\mu}, a_L^{\nu}] =
     \hat{\xi}_{L}(x) \hat{F}_L^{\mu\nu}(x) \hat{\xi}_L^{\dagger}(x) \nonumber \\
    & &  F_R^{\mu\nu}(x) = \partial^{\mu}a_R^{\nu} - \partial^{\nu}a_R^{\mu} -i [ a_R^{\mu}, a_R^{\nu}] =
     \hat{\xi}_{R}(x) \hat{F}_R^{\mu\nu}(x) \hat{\xi}_R^{\dagger}(x)
   \end{eqnarray}
   All the field strength transforms covariantly
   \begin{eqnarray}
    & &  F_L^{\mu\nu}(x)\rightarrow g_L  F_L^{\mu\nu} g_L^{\dagger}; \qquad  F_R^{\mu\nu}(x)
    \rightarrow g_R  F_R^{\mu\nu} g_R^{\dagger}\ ,\nonumber \\
     & & \hat{F}_L^{\mu\nu}(x) \rightarrow G_L  \hat{F}_L^{\mu\nu} G_L^{\dagger}; \qquad
     \hat{F}_R^{\mu\nu}(x) \rightarrow G_R  \hat{F}_R^{\mu\nu} G_R^{\dagger}
   \end{eqnarray}
   We can also construct the gauge fields
   \begin{eqnarray}
   & &  -\hat{a}_L(x) \equiv \xi_{M}(x) iD \xi_M^{\dagger}(x)=
    \xi_{M}(x) (i\partial + \hat{A}_R(x) ) \xi_M^{\dagger}(x) - \hat{A}_L(x) \\
   & &  -\hat{a}_R(x) \equiv \xi_{M}^{\dagger}(x) iD \xi_M(x)=
    \xi_{M}^{\dagger}(x) (i\partial + \hat{A}_L(x) ) \xi_M(x) - \hat{A}_R(x)
    = \xi_M^{\dagger}(x) \hat{a}_L(x) \xi_{M}(x)
   \end{eqnarray}
   which transform covariantly  under the local chiral symmetry transformation
   \begin{eqnarray}
     \hat{a}_L(x) \rightarrow G_{L}(x)\ \hat{a}_L(x)\  G_L^{\dagger}(x), \qquad
     \hat{a}_R(x) \rightarrow G_{R}(x)\ \hat{a}_R(x)\  G_R^{\dagger}(x)
   \end{eqnarray}

     With the above definitions and analysis, one can establish a more general chiral Lagrangian which
     is invaraint under the transformations of global chiral symmetry U(3)$_L$ $\times$ U(3)$_R$
     with the local chiral symmetry $\hat{U}$(3)$_L$ $\times$ $\hat{U}$(3)$_R$ appearing as a hidden symmetry.
     The more general Lagrangian has the following form
     \begin{eqnarray}
      L = & & L_a + L_b + L_c + L_d + L_k + L_f     \\
      L_a = & & a(f_{\pi}^2/16) Tr \left(a_{L\mu} + Ua_{R\mu}U^{\dagger}\right)^2 \nonumber \\
      = & & -a(f_{\pi}^2/16) Tr \left(\hat{\xi}_{L} D_{\mu} \hat{\xi}_L^{\dagger}  +
      (\hat{\xi}_{L}\xi_M) (D_{\mu} \hat{\xi}_R^{\dagger}\ \hat{\xi}_{R}) (\xi_M^{\dagger}
       \hat{\xi}_{L}^{\dagger})\right)^2   \\
      L_b = & & b(f_{\pi}^2/4) Tr \left(a_{L\mu} - Ua_{R\mu}U^{\dagger}\right)^2 \nonumber \\
       = & & -a(f_{\pi}^2/16) Tr \left(\hat{\xi}_{L} D_{\mu} \hat{\xi}_L^{\dagger}  -
      (\hat{\xi}_{L} \xi_M) (D_{\mu} \hat{\xi}_R^{\dagger}\ \hat{\xi}_{R})
      (\hat{\xi}_{L}\xi_M)^{\dagger}\right)^2   \\
       L_c = & & c(f_{\pi}^2/16) Tr (\hat{\xi}_L\hat{a}_{L\mu}\hat{\xi}_L^{\dagger})^2
        = c(f_{\pi}^2/16) Tr (\hat{a}_{R\mu})^2  =  -c(f_{\pi}^2/16)
      Tr \left(\xi_{M} D_{\mu} \xi_M^{\dagger} \right)^2 \\
      L_d = & & d(f_{\pi}^2/16) Tr (a_{L\mu} - Ua_{R\mu}U^{\dagger} - \hat{\xi}_L\hat{a}_{L\mu}\hat{\xi}_L^{\dagger})^2
      \nonumber \\
       = & & -d(f_{\pi}^2/16) Tr \left( \hat{\xi}_{L} D_{\mu} \hat{\xi}_L^{\dagger}  -
      (\hat{\xi}_{L} \xi_M) (D_{\mu} \hat{\xi}_R^{\dagger}\  \hat{\xi}_{R})
      (\hat{\xi}_{L}\xi_M)^{\dagger} - \hat{\xi}_{L} (\xi_{M} D_{\mu} \xi_M^{\dagger})
       \hat{\xi}_{L}^{\dagger} \right)^2 \\
       L_k = & &  -\frac{1}{4g^2_G}  Tr \left( F_L^{\mu\nu} F_{L\mu\nu} +  F_R^{\mu\nu} F_{R\mu\nu} \right) =
       -\frac{1}{16g^2_G}  Tr \left( \hat{F}_L^{\mu\nu} \hat{F}_{L\mu\nu} +  \hat{F}_R^{\mu\nu} \hat{F}_{R\mu\nu} \right) \\
       L_f = & & \alpha (1/12g_G^2) Tr \left( \hat{\xi}_L (D_{\mu}\hat{a}_{L\nu}) (D^{\mu}\hat{a}_{L}^{\nu})
       \hat{\xi}_L^{\dagger} \right) \nonumber \\
       & & + \beta (1/12g_G^2) Tr \left( \hat{\xi}_L \hat{a}_{L\mu} \hat{a}_{L\nu} \hat{a}_{L}^{\mu} \hat{a}_{L}^{\nu}
         \hat{\xi}_L^{\dagger} \right) \nonumber \\
         & & + \gamma (1/12g_G^2) Tr \left( \hat{\xi}_L \hat{a}_{L\mu} \hat{a}_{L}^{\mu}\hat{\xi}_L^{\dagger} \right)^2
         \nonumber \\
       & & + \alpha_1 (-i/g_G^2) Tr \left( a_{L\mu}a_{L\nu} F_L^{\mu\nu} +
       a_{R\mu}a_{R\nu} F_R^{\mu\nu} \right) \nonumber \\
       & & + \alpha_2 (-i/g_G^2) Tr \left( U^{\dagger} a_{L\mu}a_{L\nu} U F_R^{\mu\nu} +
       U a_{R\mu}a_{R\nu} U^{\dagger} F_L^{\mu\nu} \right) \nonumber  \\
       & & + \alpha_3 (+i/2g_G^2) Tr \left( a_{L\mu} U a_{R\nu} U^{\dagger}F_L^{\mu\nu} +
       a_{R\mu}U^{\dagger} a_{L\nu} U  F_R^{\mu\nu} \right) + H.c. \nonumber \\
       & & + \alpha_4 (-i/4g_G^2) Tr \left( \hat{\xi}_L \hat{a}_{L\mu} \hat{a}_{L\nu} \hat{\xi}^{\dagger}_L F_L^{\mu\nu}  +
       \hat{\xi}_R \xi_M^{\dagger}\hat{a}_{L\mu}\hat{a}_{L\nu} \xi_M \hat{\xi}_R^{\dagger} F_R^{\mu\nu}  \right)
        \nonumber \\
        & & + \alpha_5 (+i/4g_G^2) Tr \left( a_{L\mu} \hat{\xi}_L  \hat{a}_{L\nu}
       \hat{\xi}^{\dagger}_L F_L^{\mu\nu}  - a_{R\mu} \hat{\xi}_R  \hat{a}_{R\nu}
       \hat{\xi}^{\dagger}_R F_R^{\mu\nu} \right) + H.c.\nonumber \\
       & & + \alpha_6 (-i/4g_G^2) Tr \left( U a_{R\mu} U^{\dagger} \hat{\xi}_L  \hat{a}_{L\nu}
       \hat{\xi}^{\dagger}_L F_L^{\mu\nu}  -  U^{\dagger}   a_{L\mu} U \hat{\xi}_R  \hat{a}_{R\nu}
       \hat{\xi}^{\dagger}_R F_R^{\mu\nu} \right) + H.c.
     \end{eqnarray}
     For comparison, the coupling constants are taken in terms of
     the same notation as the ones in ref.\cite{BFY,BKY,UM}, except three additional interaction terms of
     the dimension four, which have been missed in ref.\cite{BFY,BKY,UM} and will be found to be very important
     for understanding the $\rho\pi\pi$ coupling $g_{\rho\pi\pi}$, and the decay rates of $a_1 \rightarrow \rho \pi$ and
     $a_1 \rightarrow \gamma \pi$. It is seen that there are fourteen unknown coupling
     constants: $a$, $b$, $c$, $d$, $g_G$, $\alpha$, $\beta$, $\gamma$, and $\alpha_i$
     ($i=1,\cdots, 6$). In general, they need to be determined via experiment processes. Actually, the success of
     current algebra helps to fix some of the couplings. It has been shown\cite{BFY,BKY,UM} that
     the following choice of the parameters seem to be consistent with the low energy phenomena and current algebra
     \begin{eqnarray}
      & & a = b = c = 2, \qquad d=0      \\
      & & \alpha_1 = \alpha_2 =\alpha_3 = 0, \qquad -\alpha_4 = \alpha_5 = \alpha_6 = 1  \\
      & & \alpha, \  \beta, \  \gamma- \  \mbox{missed}, \quad or \quad \alpha=\beta=\gamma =0
     \end{eqnarray}
    Note that the values of this set of parameters were phenomenologically suggested without including the three
    terms $\alpha$, $\beta$ and $\gamma$.  The three additional terms corresponding to the couplings
    $\alpha$, $\beta$ and $\gamma$ are introduced at first time in this paper from hidden local symmetry approach,
    we will discuss their values below and will also comment on the values of the
    other parameters based on the effective chiral Lagrangian derived from the chiral symmetry of chiral quarks and
    bound state solutions of nonperturbative QCD.

       It is clear that the physical observables should be independent
     of the Hidden symmetry, which means that we can choose any appropriate gauge for the local hidden symmetry
     to obtain the effective chiral Lagragian to describe the low energy dynamics of QCD. For convenient, we make the
     following choice for the gauge transformations $G_{L,R}(x)$, so that
   \begin{eqnarray}
    & & \xi_M(x)\rightarrow G_L(x)\xi_M(x)G_R^{\dagger}(x) = 1,    \\
    & & \hat{\xi}_L(x) \rightarrow \hat{\xi}_L(x)G_L^{\dagger}(x)= \xi_L(x) = \xi(x) = e^{i\Pi(x)/f_{\pi}} \nonumber \\
    & & \hat{\xi}_R(x) \rightarrow \hat{\xi}_R(x)G_R^{\dagger}(x)=\xi_R(x) = \xi^{\dagger}(x) = e^{-i\Pi(x)/f_{\pi}} \\
    & & U(x) = \xi_L(x) \xi_R^{\dagger}(x) = \xi^2(x) =  e^{i2\Pi(x)/f_{\pi}}
    \end{eqnarray}
    where $\Pi(x) = \Pi^a \lambda^a $ is the nonet Goldstone-like pseudoscalars. In this convention, $f_{\pi} = 186$ MeV.
    With this choice of gauge, we have
    \begin{eqnarray}
   \hat{a}_R(x) & = & - \hat{a}_L(x) = \hat{A}_R(x) - \hat{A}_L(x)  \nonumber \\
   & = & \xi_L^{\dagger} (-iDU) \xi_R  =  \xi_R^{\dagger} (iDU^{\dagger}) \xi_L
   \end{eqnarray}
   with the covariant derivative $iDU$ being
    \begin{eqnarray}
   iDU = i\partial U + a_L U - Ua_R
   \end{eqnarray}
   It is seen that the above choice of gauge conditions is a kind of unitary gauge corresponding to
   the broken down of the hidden local chiral symmetry.
   
   It will also be useful to decompose the chiral gauge fields $\hat{A}_L$ and $\hat{A}_R$ into two parts
   \begin{eqnarray}
    \hat{A}_L(x) \equiv A_{L}(x) + L_{\xi}(x), \qquad
     \hat{A}_R(x) \equiv A_{R}(x) + R_{\xi}(x)
   \end{eqnarray}
   with $A_{L}(x)$ and $A_{R}(x)$ being the covariant parts associated with
   the gauge bosons $a_L(x)$ and $a_R(x)$, while $L_{\xi}(x)$ and
   $R_{\xi}(x)$ are the pure gauge part associated with the Goldstone-like pseudoscalars
   contained in the nonlinear chiral fields $\hat{\xi}_L(x)$ and $\hat{\xi}_L(x)$
   \begin{eqnarray}
    & &  A_{L}(x) = \xi_{L}^{\dagger}(x)\ a_L(x)\ \xi_L(x) \equiv V(x)-A(x), \\
    & &  L_{\xi}(x)=\xi_{L}^{\dagger}(x)i\partial \xi_L(x) \equiv V_{\xi}(x)-A_{\xi}(x), \\
    & &  A_{R}(x) = \xi_{R}^{\dagger}(x)\ a_R(x)\ \xi_R(x)\equiv V(x) + A(x), \\
    & &  R_{\xi}(x)=\xi_{R}^{\dagger}(x)i\partial \xi_R(x)\equiv V_{\xi}(x)+A_{\xi}(x)
   \end{eqnarray}
   Explicitly, one sees that
   \begin{eqnarray}
    2A_{\xi} = R_{\xi}(x)- L_{\xi}(x) = \xi_L^{\dagger} (-i\partial U )  \xi_R
   \end{eqnarray}
    With the above gauge choice, we arrive at the following effective chiral Lagrangian possessing only global
    chiral symmetry U(3)$_L$ $\times$ U(3)$_R$

  \begin{eqnarray}
      L_t = & & L_a + L_b + L_c + L_d \nonumber \\
      & & =(a+b) (f_{\pi}^2/16) Tr \left( a_{L\mu}^2 + a^2_{R\mu}\right) + 2(a-b)(f_{\pi}^2/16) Tr
      \left( a_{L\mu}Ua_{R}^{\mu}U^{\dagger} \right)  \nonumber \\
      & & +c(f_{\pi}^2/16) Tr \left(DU DU^{\dagger}\right) +
      d(f_{\pi}^2/16) Tr (a_{L\mu} - Ua_{R\mu}U^{\dagger} - iDU\ U^{\dagger})^2  \\
      L_f = & & \alpha (1/12g_G^2) Tr \left( D_{\mu}D_{\nu}U\ D^{\mu}D^{\nu}U^{\dagger} \right) \nonumber  \\
      & & +  \beta (1/12g_G^2) Tr \left( D_{\mu}UD_{\nu}U^{\dagger} D^{\mu}UD^{\nu}U^{\dagger} \right) \nonumber \\
      & & +  \gamma (1/12g_G^2) Tr \left( D_{\mu}UD^{\mu}U^{\dagger} D_{\nu}UD^{\nu}U^{\dagger} \right) \nonumber \\
      & & + \alpha_1 (-i/g_G^2) Tr \left( a_{L\mu}a_{L\nu} F_L^{\mu\nu} +
       a_{R\mu}a_{R\nu} F_R^{\mu\nu} \right) \nonumber \\
       & & + \alpha_2 (-i/g_G^2) Tr \left(  a_{L\mu}a_{L\nu} U F_R^{\mu\nu}U^{\dagger} +
        a_{R\mu}a_{R\nu} U^{\dagger} F_L^{\mu\nu}U \right) \nonumber  \\
       & & + \alpha_3 (+i/2g_G^2) Tr \left( a_{L\mu} U a_{R\nu} U^{\dagger}F_L^{\mu\nu} +
       a_{R\mu}U^{\dagger} a_{L\nu} U  F_R^{\mu\nu} \right) + H.c. \nonumber \\
       & & + \alpha_4 (-i/4g_G^2) Tr \left( D_{\mu}UD_{\nu}U^{\dagger} F_L^{\mu\nu}  +
       D_{\mu}U^{\dagger}D_{\nu}U F_R^{\mu\nu}  \right)
        \nonumber \\
        & & + \alpha_5 (-i/4g_G^2) Tr \left( a_{L\mu} iD_{\nu}U\ U^{\dagger} F_L^{\mu\nu}
         - a_{R\mu} iD_{\nu}U^{\dagger}\ U  F_R^{\mu\nu} \right) + H.c.\nonumber \\
       & & + \alpha_6 (-i/4g_G^2) Tr \left( U a_{R\mu} iD_{\nu}U^{\dagger}F_L^{\mu\nu}  -
      U^{\dagger}  a_{R\mu} iD_{\nu}U\ F_R^{\mu\nu}  \right) + H.c.
       \end{eqnarray}
  We will see below that it is this form of the effective chiral Lagrangian that enables us to compare it with
  the one derived from ECT. This is because they possess the same global chiral
 symmetry U(3)$_L$ $\times$ U(3)$_R$ in the limit of zero light quark masses $m_q\rightarrow 0$.
 It then allows us to fix the fourteen parameters in terms of two parameters introduced in the
 effective chiral theory of mesons in the large $N_c$ approach.

 \section{Chiral Lagrangian from Large $N_C$ QCD at Low Energy}

 Let us begin with the QCD Lagrangian with only containing light quarks

 \begin{eqnarray}
 L_{QCD}^q = \bar{q}\gamma^{\mu}(i\partial_{\mu} + g_s G_{\mu}^a T^a)q -\bar{q}M_q q - \frac{1}{2} Tr G_{\mu\nu}G^{\mu\nu}
 \end{eqnarray}
 where $q =(u, d, s)$ denote three light quarks. $G_{\mu}^a$ are the gluon fields with SU(3) gauge symmetry
 and $g_s$ is the running coupling constant. $M_q$ is the light quark mass matrix $M_q = diag.(m_u, m_d, m_s)$.
 In the limit $M_q\rightarrow 0$, it is seen that the Lagrangian has global $U(3)_L \times U(3)_R$ symmetry.
 It is known that the pseudoscalars, vector and axial-vector mesons (or chiral spin-1 bosons) all are bound states
 of quark and anti-quark due to strong interactions of gluons at low energy. The global chiral symmetry will be
 broken down spontaneously via the dynamical mechanism of quark condensate due to attractive gluon interactions at
 low energy. When the confining length scale of quark and anti-quark in mesons is smaller than the scattering length
 among the mesons, one may treat the mesons as point-like quantum fields of composite quarks. In the classical level,
 it is supposed to have the following correspondings between the mesons and the quark currents
 \begin{eqnarray}
  \Pi^i(x) \sim G(q, \bar{q}) \bar{q}(x)i\gamma_5\lambda^i q(x); \qquad v^i_{\mu}(x) \sim \bar{q}(x)\gamma_{\mu}\lambda^i q(x);
  \qquad   a^i_{\mu}(x) \sim \bar{q}(x)\gamma_{\mu}\gamma_5\lambda^i q(x)
 \end{eqnarray}
To describe such phenomena of the low energy dynamics of QCD, one can extend the QCD Lagrangian
at low energy to the following form with including the meson fields as bound states
 \begin{eqnarray}
 \hat{L}_{QCD}^q & = & \bar{q}\gamma^{\mu}(i\partial_{\mu} + g_s G_{\mu}^a T^a)q -\bar{q}M_q q
 - \frac{1}{2} Tr G_{\mu\nu}G^{\mu\nu} \nonumber \\
 & + & \bar{q}(x)\left(\gamma^{\mu}v^i_{\mu}(x) \lambda^i + \gamma_{\mu}\gamma_5 a^i_{\mu}(x)\lambda^i
 + mU(x)P_+ + mU^{\dagger}P_{-} \right) q(x) \nonumber \\
 &+& \frac{m_o^2}{4} Tr (v^2(x) + a^2(x))
 \end{eqnarray}
 with $P_{\pm} = (1\pm \gamma_5)/2$. The above Lagrangian remains invaraint under transformations of the global chiral
 symmetry $U(3)_L \times U(3)_R$  in the limit $m_q \rightarrow 0$,
 \begin{eqnarray}
  & & q_L(x)\equiv P_+ q(x)\rightarrow g_L q_{L}(x), \quad  q_R(x)\equiv P_- q(x)\rightarrow g_R q_{R}(x);
  \quad U(x) \rightarrow
  g_L U(x) g_{R}^{\dagger}, \nonumber \\
  & &  a_L(x) \equiv v(x) - a(x) \rightarrow g_L a_L(x) g_L^{\dagger}, \quad
  a_R(x) \equiv v(x) + a(x) \rightarrow g_R a_R(x) g_R^{\dagger}
 \end{eqnarray}
 Note that the pseudoscalars are treated as the nonlinear fields. Based on the above considerations,
 mesons are regarded as the bound state solutions of $QCD$, namely they
 are not independent degrees of freedom. Therefore, there are no kinetic terms for meson fields.
 The kinetic terms of meson fields are generated dynamically from quark loops.

  The effective Lagrangian of mesons is obtained by integrating out quark fields in Eq.(42)
 (which is equivalent to calculate the Feynman diagrams of quark loops).
 Using the method of path integral, the effective Lagrangian of mesons is evaluated via
\begin{equation}
exp \{i\int d^{4}x{\cal L}^{M}\}=\int[dq][d\bar{q}]
exp \{i\int d^{4}x{\cal L}\}.
\end{equation}
The functional integral of right hand side is known as the determination of the Dirac operator
\begin{equation}
 \int[dq][d\bar{q}]
exp \{i\int d^{4}x{\cal L}\} = \det({\cal D}).
\end{equation}
To obtain the effective Lagrangian, it will be useful to go to Euclidean space and to define the Hermitian
operator
\begin{equation}
{\cal L}^{M}_{E}=\ln \det{\cal D} = \frac{1}{2}[\ln \det{\cal D} + \ln \det{\cal D}^{\dagger} ]
+ \frac{1}{2}[\ln \det{\cal D} - \ln \det{\cal D}^{\dagger} ] \equiv {\cal L}_{Re}^M + {\cal L}_{Im}^M
\end{equation}
with
\begin{eqnarray}
{\cal L}_{Re}^M & = & \frac{1}{2}\ln \det ({\cal D}{\cal D}^{\dagger}) \equiv \frac{1}{2}\ln \det\Delta_E  \nonumber \\
{\cal L}_{Im}^M & = & \frac{1}{2}\ln \det ({\cal D} /{\cal D}^{\dagger}) \equiv \frac{1}{2}\ln \det\Theta_E
\end{eqnarray}
where
\begin{eqnarray}
& & {\cal D}=\gamma\cdot \partial-i\gamma \cdot v(x)-i\gamma\cdot a(x) \gamma_{5} + m u(x) \, \nonumber \\
& & {\cal D}^{\dagger}=-\gamma\cdot \partial + i\gamma \cdot v(x)-i\gamma\cdot a(x) \gamma_{5} + m \hat{u}(x)
\end{eqnarray}
with $u(x) = P_+ U(x) + P_- U^{\dagger}(x)$ and $\hat{u}(x)=  P_- U(x) + P_+ U^{\dagger}(x)$. Applying the following
useful relations
\begin{eqnarray}
& & \ln \det \Delta = Tr \ln \Delta \\
& & \ln \Delta  = -\int_{\tau_0}^{\infty}\frac{d\tau}{\tau} e^{-\tau \Delta} -\left( \ln \tau_0 + \gamma +
 \sum_{n=1}^{\infty} \frac{(-\tau_0 \Delta)^n}{n\cdot n !} \right)
\end{eqnarray}
The effective Lagrangian can be obtained by using the Schwinger's proper time regularization approach\cite{JS,Ball}
with subtracting the divergence at $\tau_0 =0$
 \begin{eqnarray}
{\cal L}_{Re}^M  ={1\over 2}
\int d^{4}x\frac{d^{4}p}{(2\pi)^{4}}Tr\int^{\infty}
_{0}{d\tau\over \tau} \left( e^{-\tau \Delta_E^p }
-e^{-\tau\Delta_{0}} \right)
\end{eqnarray}
with
\begin{eqnarray}
& & \Delta_E^p = (\gamma^{\mu} D^{\dagger}_{\mu})(\gamma^{\nu} D_{\nu}) + 2ip^{\mu} D_{\mu} + m\gamma^{\mu}
D_{\mu}U(x) P_+ + m\gamma^{\mu} D_{\mu}U^{\dagger}(x) P_-  + \Delta_0 \nonumber \\
& & \Delta_0 = p^{2} + m^2
\end{eqnarray}
and
\begin{eqnarray}
 D_{\mu} = \partial_{\mu} -iv_{\mu}(x) - i a_{\mu}(x) \gamma_5 \ , \quad
 D_{\mu}^{\dagger} = \partial_{\mu} -iv_{\mu}(x) + i a_{\mu}(x) \gamma_5
\end{eqnarray}

To ensure the gauge invariance, the dimensional regularization should be adopted for the momentum integral. After
 integrating over $\tau$, the effective Lagrangian reads
\begin{eqnarray}
{\cal L}_{Re}^{M}={1\over 2}\int d^{D}x\frac
{d^{D}p}{(2\pi)^{D}}\sum^{\infty}_{n=1}{1\over n}\frac{1}{
(p^{2}+m^{2})^{n}} \ Tr \left( \Delta_E^p - \Delta_0 \right)^n
\end{eqnarray}

 Up to the fourth order in covariant derivatives,  the effective Lagrangian
 in Minkowski space has the following form
\begin{eqnarray}
{\cal L}_{Re}^M & = & \frac{F^2}{16} Tr (D_{\mu}UD^{\mu}U^{\dagger})
 -\frac{g^2}{8}\ Tr \left( v_{\mu\nu}v^{\mu\nu}
 + a_{\mu\nu}a^{\mu\nu}\right) \nonumber \\
 & &+\frac{N_{c}}{6(4\pi)^{2}}TrD_{\mu}D_{\nu}UD^{\mu}D^{\nu}U^{\dagger}
\nonumber \\
 & &-{i\over 2}\frac{N_{c}}{(4\pi)^{2}}Tr\{D_{\mu}UD_{\nu}U^{\dagger}+
D_{\mu}U^{\dagger}D_{\nu}U\}v^{\mu\nu}\nonumber \\
 & &-{i\over 2}\frac{N_{c}}{(4\pi)^{2}}Tr\{D_{\mu}U^{\dagger}D_{\nu}U-D_{\mu}
UD_{\nu}U^{\dagger}\}a^{\mu\nu} \nonumber \\
 & &-\frac{N_{c}}{12(4\pi)^{2}}Tr\{
2\ D_{\mu}UD^{\mu}U^{\dagger}D_{\nu}UD^{\nu}U^{\dagger}
-D_{\mu}UD_{\nu}U^{\dagger}D^{\mu}UD^{\nu}U^{\dagger}\} \nonumber \\
 & &+{1\over 4}m^{2}_{0} Tr \left( v_{\mu}
v^{\mu}+a_{\mu}a^{\mu} \right),
\end{eqnarray}
where
\begin{eqnarray*}
& & D_{\mu}U=\partial_{\mu}U-i[v_{\mu}, U]+i\{a_{\mu}, U\},\\
& & D_{\mu}U^{\dagger}=\partial_{\mu}U^{\dagger}-i[v_{\mu}, U^{\dagger}]-
 i\{a_{\mu}, U^{\dagger}\},\\
& & v_{\mu\nu}=\partial_{\mu}v_{\nu}-\partial_{\nu}v_{\mu}
-i[v_{\mu}, v_{\nu}]-i[a_{\mu}, a_{\nu}],\\
& & a_{\mu\nu}=\partial_{\mu}a_{\nu}-\partial_{\nu}a_{\mu}
-i[a_{\mu}, v_{\nu}]-i[v_{\mu}, a_{\nu}],\\
& & D_{\nu}D_{\mu}U=\partial_{\nu}(D_{\mu}U)-i[v_{\nu}, D_{\mu}U]
+i\{a_{\nu}, D_{\mu}U\},\\
& & D_{\nu}D_{\mu}U^{\dagger}=\partial_{\nu}(D_{\mu}U^{\dagger})
-i[v_{\nu}, D_{\mu}U^{\dagger}]
-i\{a_{\nu}, D_{\mu}U^{\dagger}\}.
\end{eqnarray*}
The two parameters entering the effective Lagrangian are given by
\begin{eqnarray}
& & F^2 = 16\frac{N_{c}}{(4\pi)^{2}}m^{2}{D\over 4}\Gamma(2-{D\over 2}) \left(\frac{4\pi\mu^2_I}{m^2}\right)^{2-D/2} \\
& & g^2 = {8\over 3}\frac{N_{c}}{(4\pi)^{2}}{D\over 4}\Gamma(2-{D\over 2}) \left(\frac{4\pi\mu^2_I}{m^2}\right)^{2-D/2}
= \frac{F^2}{6m^2}
\end{eqnarray}
with $\mu_I$ being the infrared cutoff energy scale.

 The above effective chiral Lagrangian has been shown to provide a successful
 description on many processes of low energy phenomena\cite{BAL1,BAL2}.

 \section{14-Parameters in Chiral Lagrangian of Hidden Symmetry}

  So far we have presented, in terms of two approaches, the effective chiral Lagrangian
  with including the vector and axial-vector mesons.
  One is based on  the explicit global chiral symmetry $U(3)_L \times U(3)_R$  and hidden local  chiral symmetry
  $\hat{U}$(3)$_L$ $\times$ $\hat{U}$(3)$_R$, another based on the global chiral symmetry of chiral quarks
  and bound state solutions of nonperturbative QCD at low energy. With the approapriate gauge choice of the
  hidden local chiral symmetry as shown in eqs.(38-39) in the section II, it allows us to fix the fourteen
  parameters introduced in the hidden local chiral symmetry approach. It is not difficult
  to check that the parameters are fixed to be
  \begin{eqnarray}
  & & a = b = \frac{m_0^2}{f_{\pi}^2}=\frac{g^2 m_{\rho}^2}{f_{\pi}^2}, \qquad
  c = \frac{6g^2 m^2}{f_{\pi}^2} = \frac{F^2}{f_{\pi}^2}, \qquad d =0 \\
  & & \alpha = 2\beta = - \gamma =  \frac{N_c}{2(\pi g)^2}, \qquad g_G^2 = \frac{4}{g^2}    \\
  & & \alpha_1=  \alpha_2 =  \alpha_3 = \alpha_5 = \alpha_6 = 0 ,  \qquad \alpha_4 =  \frac{N_c}{2(\pi g)^2} = \alpha
  \end{eqnarray}
  To define the physical meson states in the mass eigenstate, one needs to normalize the kinetic terms and
  redefine the pseudoscalars and axial-vectors due to the mixing term $a^{\mu}(x)\partial_{\mu}\Pi(x)$, which leads to
  \begin{eqnarray}
  & & f_{\pi}^2 = F^2 \left(1- \frac{2c}{a+b+2c}\right) =
   F^2 \left(1- \frac{6m^{2}}{ m_{\rho}^2 +6m^2}\right)  \\
  & & m_{\rho}^2 = m_o^2/g^2
 \end{eqnarray}
 Comparing the above set of parameters with the one fixed via the low energy phenomena (eqs.(24-26)),
 we come to the following observations:
 \begin{itemize}
 \item For the terms in the effective Lagrangian up to the dimension-two of covariant derivative,
 both effective chiral theory approach
 and hidden symmetry approach provide a consistent determination for the four parameters.
 It is of interest to note that once the vector meson mass is dynamically generated and takes the value
 \begin{equation}
  m_{\rho}^2 = 6m^2
 \end{equation}
 one has from eqs.(58) and (61)
 \[F^2 = 2f_{\pi}^2, \qquad  a=b=c = 2 \]
 which agrees well with the one obtained from current algebra and phenomenological analysis (eq.(24))
 in the hidden symmetry approach\cite{BFY,BKY,UM}.
 \item For the terms with dimension four of covariant derivative, it is noticed that:
 (i) there are in general ten terms rather than seven terms
 in the usual effective Lagrangian of hidden symmtry in the literature\cite{BFY,BKY,UM}, three additional new terms
 (i.e., $\alpha$, $\beta$ and $\gamma$) are necessity in our present more general construction on the effective
 Lagrangian via the hidden symmetry approach. Of particular, these three terms are found to be nonzero when
 comparing to the effective chiral Lagrangian derived from ECT;
 (ii) Even for the usual six terms with coupling constants $\alpha_i$, $i=1,\cdots, 6$, three of the terms,
 $\alpha_4$, $\alpha_5$ and $\alpha_6 $,  turn out to have different behavior
 when comparing their values yielded from phenomenological analysis in
 the literatures\cite{BFY,BKY,UM} with the ones determined from the effective chiral theory.
 \item  The values $-\alpha_4 = \alpha_5 = \alpha_6 = 1$ have been taken in the literature\cite{BFY,BKY,UM} to
 accomodate the $\rho-$dominance for $a_1\rightarrow \gamma \pi$ decay and to cancel the
 strong momentum dependence of the
 coupling $f_{\rho\pi\pi}$ in the absence of the $a_1-$ meson. While in the effective chiral theory, it is seen that
 $\alpha_4$ is positive with value $\alpha_4 = N_c /(2(\pi g)^2)$, and $\alpha_5 = \alpha_6 = 0$.
 \item It is natural to ask why the values $\alpha_4$, $\alpha_5$ and $\alpha_6 $ extracted from two cases
 are so different, and how the cancellation of strong momentum dependence of the coupling $f_{\rho\pi\pi}$ and
 the $\rho-$ dominance in $a_1\rightarrow \gamma \pi$ decay can be accomodated in the case with positive value of
 $\alpha_4$ and zero values of $\alpha_5$ and $\alpha_6$. The answer is attributed to the three additional new terms,
 in our present more general construction via hidden symmetry approach. They are found to be nonzero from
 the effective chiral theory and their values are determined from the effective chiral theory to be
 $\alpha = - \gamma = 2\beta = \alpha_4 =  N_c/(2(\pi g)^2)$.  With these values, it can be shown that
 the strong momentum dependence of $f_{\rho\pi\pi}$ will be cancelled when $m_{\rho}^2 = 6m^2$ and
 $g=1/\pi$ due to the
 existance of the additional new terms, and the $\rho-$dominance for $a_1\rightarrow \gamma \pi$ decay
 can also be realized\cite{BAL1,BAL2}.
 \item In comparison with the chiral perturbation theory (ChPT) \cite{ChPT}, the new terms
 $\alpha$, $\beta$ and $\gamma$
 are related to the terms $L_1$, $L_2$ and $L_3$ in ChPT. Noticing the algebraic relation
 \begin{eqnarray}
 & & Tr \left( D_{\mu}UD_{\nu}U^{\dagger}D^{\mu}UD^{\nu}U^{\dagger}\right) = \frac{1}{2}
  [Tr \left( D_{\mu}UD^{\mu}U^{\dagger} \right) ]^2  \nonumber \\
  & & + Tr \left( D_{\mu}UD_{\nu}U^{\dagger} \right)\cdot Tr\left( D^{\mu}UD^{\nu}U^{\dagger}\right)
  -2 Tr \left( D_{\mu}UD^{\mu}U^{\dagger}\right)^2
 \end{eqnarray}
 One arrives at the relation $L_1 = 1/2\ L_2$.
 \item The terms $\alpha_4$ and $\alpha$ are related to the coupling constant $L_9$ in ChPT. Both the sign and extracted value
 for $\alpha_4$ in our present consideartions are consistent with the ones of $L_9$ from the phenomenology well descriped by ChPT,
 while the previous results for $\alpha_4$ given in literature\cite{BFY,BKY,UM} seem to be conflict with the extracted value of
 $L_9$ in ChPT.
\end{itemize}

  It is then not difficult to show that the more general effective Lagrangian (eqs.(38-39))
  constructed via the approach of global chiral symmetry and hidden local chiral symmetry with an appropriate
  gauge choice should be consistent with any other effective chiral Lagarangian in the chiral limit.
  The fourteen parameters in the effective Lagrangian up to the
  dimension-four of the meson fields can be extracted from the effective chiral theory.

\section{ Effective Chiral Lagrangian and Low Energy Behavior}

  A consistent effective chiral Lagrangian should reproduce the low energy behaviour which have been tested
  by experiments. Let us check the Vector-Pseudoscalar-Pseudoscalar vertex. As an example, we may first work out
  the $\rho\pi\pi$ coupling $f_{\rho\pi\pi}$ which is defined as
  \begin{equation}
  {\cal L}_{\rho\pi\pi} = f_{\rho\pi\pi}\epsilon_{ijk}\rho_i^{\mu}\pi_j\partial_{\mu} \pi_k
  \end{equation}
  From the general effective chiral Lagrangian eqs.(38) and (39), it is not difficult to find that
  \begin{eqnarray}
  f_{\rho\pi\pi} & = & g_G \{\ 1+ \frac{2m_{\rho}^2}{g_G^2f_{\pi}^2} \  [ \ (\alpha_4-\alpha/3)\left(1-\frac{2c}{a+b + 2c}\right)^2
   -\left(\frac{2c}{a+b + 2c}\right)^2 \nonumber \\
   & + &  (\alpha_5 + \alpha_6) \left(\frac{2c}{a+b + 2c}\right)\left(1-\frac{2c}{a+b + 2c}\right) \  ] \  \}
  \end{eqnarray}
  Note that the coupling receives contribution from the additional new term $\alpha$.
   It is seen that when the parameters take the values chosen  from the phenomenological analysis in the
  literature\cite{BFY,BKY,UM}, i.e., $a=b=c=2$, $-\alpha_4 = \alpha_5 = \alpha_6 =1$ and $\alpha =0$
  (also see eqs.(24-26)), one has
   \begin{eqnarray}
  f_{\rho\pi\pi} = g_G =2/g
  \end{eqnarray}
  where the second term in the curled bracket of eq.(66) vanishes due to cancellations from various contributions.
  Alternatively, when taking the values of parameters determined from the effective
  chiral theory, see eqs. (58-60), one has\cite{BAL1}
  \begin{eqnarray}
  f_{\rho\pi\pi} = \frac{2}{g} \{\  1 + \frac{m_{\rho}^2}{2\pi^2f_{\pi}^2}\left[ \frac{N_c}{3} \left(1-\frac{6m^2}{m_{\rho}^2 +6m^2}\right)^2
   - \pi^2g^2\left(\frac{6m^2}{m_{\rho}^2 + 6m^2}\right)^2 \right]  \  \}
  \end{eqnarray}
  It is seen that only for a specific choice $g = 1/\pi$ and $m_{\rho}^2 = 6m^2$, one yields with $N_c = 3$ that $ f_{\rho\pi\pi} =2/g$.
  It can be shown that with the parameters fixed from the effective chiral theory, the effective
  chiral Lagrangian can also lead to a consistent prediction on $\Gamma (a_1 \rightarrow \rho \pi)$
  and  $\Gamma (a_1 \rightarrow \gamma \pi)$. The numerical
  predictions were found\cite{BAL1} to be $\Gamma (a_1 \rightarrow \rho \pi) \simeq 326$ MeV and
  $\Gamma (a_1 \rightarrow \gamma \pi) \simeq 252$ MeV. In general, a value of the
  basic parameter $g$ closing to $1/\pi$ is found to be a consistent one. Here the term
  $\alpha$ play an important role.

  The second effect of the additional term $\alpha$ in the more general effective chiral Lagrangian is that
  the Weinberg's sum rule $g_{a}^2 = g_{\rho}^2$ will be modified to be
  \begin{equation}
  g_{a}^2 = g_{\rho}^2 \left(1 - \frac{\alpha}{3} \right) =  g_{\rho}^2 \left(1 - \frac{N_c}{6\pi^2g^2} \right)
  \end{equation}
  In the second equility, the parameter $\alpha$ has been taken the result fixed from the effective chiral theory.
  This modification makes the predictions for the masses of the axial vectors to be more consistent with the
  experimental data.

  Another important effect from the additional term $\alpha$ is the evaluation for the decay constants
  of the pseudoscalars.

  To be more explicit, we may use some algebraic relations and equation of motion
  \begin{equation}
  D^{\mu}\left(U^{\dagger} D_{\mu} U\right) = \frac{1}{2} \left(U^{\dagger}\chi - \chi^{\dagger} U\right)
  \end{equation}
  to reexpress the $\alpha$ term into several more familar terms, so that its effects can be easily seen.
  It is easy to check that
 \begin{eqnarray}
 D_{\mu}D_{\nu} U D^{\mu}D^{\nu} U^{\dagger} & = & \frac{1}{2}
 [ F_{L}^2  + F_{R}^2 - 2F_L U F_R U^{\dagger} ] \nonumber \\
 & + & i [ D_{\mu}U D_{\nu} U^{\dagger} F_L^{\mu\nu} + D_{\mu}U^{\dagger} D_{\nu} U F_R^{\mu\nu} ] \nonumber \\
 & + & \left(D_{\mu}U D^{\mu} U^{\dagger}\right) \left(D_{\nu}U D^{\nu} U^{\dagger}\right) \\
 & + & \frac{1}{2} D_{\mu}U D^{\mu} [ U^{\dagger} \left(U \chi^{\dagger}
 - \chi U^{\dagger} \right) ] \nonumber \\
 & + & \frac{1}{2} \left(D_{\mu}U^{\dagger} D^{\mu}  U \right) \left(U^{\dagger} \chi
 - \chi^{\dagger} U \right) \nonumber \\
 & + & \mbox{total\ derivative\ terms} \nonumber
 \end{eqnarray}
 From this explicit form, it is not difficult to understand its effects. Where the first term modifies the Weinberg's sum rule,
 the second term contributes to the $\rho\pi\pi$ coupling and the coupling constant $L_9$ in ChPT, the third term has effects
 on the coupling constant $L_3$ in ChPT and the last two terms will provide additional contributions to the decay constants of
 pseudoscalars. As a consequence, we arrive at a complete
 prediction for the couplings $L_1$, $L_2$, $L_3$ and $L_9$ at this order, which is consistent with the one extracted from phenomenology
 described by the chiral perturbation theory up to order of $p^4$. The numerical values are found to be
 \\

\begin{tabular}{|c|c|c|c|c|}
\hline
  Parameters & $10^3L_1$ & $10^3L_2$ & $10^3L_3$ & $10^3L_9$   \\ \hline
  Present    & 0.79  &  1.58 & -3.16 & 6.32 \\ \hline
  ChPT\cite{LP} & $0.4\pm 0.3$ & $1.35\pm 0.3$ & $-3.5\pm 1.1$ & $6.9\pm 0.7$ \\ \hline
 \end{tabular}
 \\
 \\

 Let us now check the known KSFR relation. From the general effective Lagrangian, the mass of $\rho$ meson reads
 $m_{\rho}^2 = a f_{\pi}^2 g_G^2/4$. Comparing to the effective chiral theory
 with $g_G^2 = 4/g^2$ and $f_{\rho\pi\pi}\simeq 2/g$, one has
  \begin{equation}
  m_{\rho}^2 = a f_{\pi}^2 g_G^2/4 = \frac{a}{4} f_{\pi}^2 \left(\frac{2}{g}\right)^2 \simeq
  \frac{a}{4} f_{\pi}^2  f^2_{\rho\pi\pi}
  \end{equation}
 Thus the known KSFR relation holds for $a\simeq 2$ which is also consistent with the vector meson dominance.

  It is seen that the more general effective chiral Lagrangian with its parameters extracted from the effective chiral theory
  can well reproduce the phenomenology of $\rho-a_1$ system.

  One may see that only from the $\rho\pi\pi$ coupling, $a_1 \rightarrow \rho\pi$ and $a_1\rightarrow \gamma \pi$
  decays, the parameters appearing in the dimension four of covariant derivative in
  the effective chiral Lagrangian constructed via hidden
 symmetry approach may not uniquely be determined. The value of the parameter $\alpha_4$
 extracted from phenomenology of $\rho-a_1$ system in the literatures\cite{BFY,BKY,UM} is conflicit with the one
 from the phenomelogy well described by the chiral perturbation theory and effective chiral theory.
 While the resulting structure and couplings from the effective chiral theory are consistent not only
 with the phenomenolgy of $\rho-a_1$ system, but also with the chiral perturbation theory. Thus the effective
 chiral theory derived from the chiral quarks and bound state solutions of nonperturbative QCD may provide a very
 useful way to extract all the parameters in terms of only two basic scales $m$ and $f_{\pi} = 186$ MeV (or coupling
 constant $g$). It is likely that the structures of the effective chiral Lagrangian for the dimension-four of
 covariant derivative given in those literatures\cite{BFY,BKY,UM}  are incomplete. As a consequence,
 the extracted coulping constants are not reliable.

 \section{Conclusions}

 The more general effective chiral Lagrangian of mesons (peusodoscalars, vectors and axial-vectors)
 has been constructed in the chiral limit by using explicit global chiral symmetry  U(3)$_L$ $\times$ U(3)$_R$
 and hidden local chiral symmetry $\hat{U}$(3)$_L$ $\times$ $\hat{U}$(3)$_R$. It has been shown that
 there are in general fourteen interacting terms up to the dimension-four of covariant derivative, which has three
 additional new terms in comparison with the effective Lagrangian given in literature. It is those three terms
 that have been found to play an important role for understanding the vector meson dominance, and
for providing consistent predictions on the decay rates of $a_1\rightarrow \gamma \pi$ and
 $a_1\rightarrow \rho \pi$, as well as for resulting a consistent effective chiral Lagragian with the chiral perturbation theory.
 It is also those three new terms that make the more general effective Lagrangian
 obtained from the symmetry considerations to be consistent with the one derived from
 the QCD Lagrangain of quarks with the global chiral symmetry and
 the bound state solutions of nonperturbative QCD at low energy with color number $N_c$.
 The latter effective chiral Lagrangian only relies on two basic parameters in addition
 to the ones in the standard model, which has
 allowed us to extract all the fourteen parameters in the more general effective Lagrangian up to dimension-four of
 covariant derivative in the chiral limit. As a consequence, it has been observed that not only the three additional new interacting terms
 introduced in this paper are necessary, but also the resulting coupling constants for other three interacting terms
 at the dimension-four have total different values in comparison with the ones given in the
 literature from hidden symmetry approach. It is likely that the structures of the effective lagrangian
 for the dimension-four terms given in the literatures\cite{BFY,BKY,UM} are incomplete, thus
 the extracted coupling constants are also not reliable.
 The effective chiral theory based on the chiral symmetry of chiral quarks and
 the bound state solutions of nonperturbative QCD at low energy and large $N_c$ shall be a more economic and reliable
 effective theory in the sense that it depends on only two basic parameters in addition to the ones
 in the standard model and can lead to a consistent prediction for the low energy phenomena of vector and axial-vector mesons.

 {\bf Acknowlodgement}

\end{document}